\documentclass[epj]{svjour}

\usepackage{epsfig}

\usepackage{graphicx,latexsym,xcolor}
\usepackage{amsmath}

\global\long\def\ket#1{\left|#1\right\rangle }

\global\long\def\bege{\begin{equation}}
\global\long\def\ende{\end{equation}}

\global\long\def\begal{\begin{align}}
\global\long\def\endal{\end{align}}

\title{Fractional-quantum-Hall-effect (FQHE) in 1D Hubbard models
\\}

\author{Ioannis Kleftogiannis\inst{1} \and Ilias Amanatidis\inst{2} }

\institute{                    
  \inst{1} Physics Division, National Center for Theoretical Sciences, Hsinchu 30013, Taiwan\\
  \inst{2} Department of Physics, Ben-Gurion University of the Negev, Beer-Sheva 84105, Israel
}

\date{\today}
\abstract{
We study the quantum self-organization of interacting particles in one-dimensional(1D) many-body systems, modeled via Hubbard chains with short-range interactions between the particles. We show the emergence of 1D states with density-wave and clustering order, related to topology, at odd denominator fillings that appear also in the fractional-quantum-Hall-effect (FQHE), which is a 2D electronic system with Coulomb interactions between the electrons and a perpendicular magnetic field. For our analysis we use an effective topological measure applied on the real space wavefunction of the system, the Euler characteristic describing the clustering of the interacting particles. The source of the observed effect is the spatial constraints imposed by the interaction between the particles. In overall, we demonstrate a simple mechanism to reproduce many of the effects appearing in the FQHE, without requiring a Coulomb interaction between the particles or the application of an external magnetic field.
}

\begin{document}

\authorrunning{I.Kleftogiannis and I. Amanatidis} 
\titlerunning{Fractional-quantum-Hall-effect (FQHE) in 1D Hubbard models}
\maketitle

Self-organization mechanisms manifesting in quantum many-body systems, can give rise to exotic phases of matter with unusual properties related to topology and quantum correlations(entanglement)\cite{spinhaldane,haldane0,AKLT,Levin,Gu,kitaev2,kitaev3,alba,alba1,hen,Hamma,Calabrese,Pollmann,amico,horodecki}. One of the earliest examples of quantum self-organization mechanisms is the fractional-quantum-Hall-effect (FQHE).
Interacting electrons confined in a 2D plane under the influence of a perpendicular magnetic field self-organize in a quantum mechanical manner, forming quantum-liquid phases of matter with several unusual properties\cite{Tsui,Laughlin,Stormer}. For instance, electron fractionalization mechanisms are present, giving rise to collective excitations that carry a fraction of the electronic charge, leading to fractional conductance at quantized plateaus with extraordinary experimental accuracy. In addition, the FQHE phases have been shown to contain topological and strong entanglement properties, thus acting as a prototype system for studying topological order\cite{Li,haldane_geometry}.

In this paper we investigate the quantum self-organization of interacting spinless particles in one dimension (1D), modeled by minimal many-body Hubbard models with short-range interactions. We demonstrate the appearance of quantum phases with density-wave and clustering order related to topology, occurring at system fillings that appear also in the FQHE. The appearance of these phases in our model does not require an external magnetic field or Coulomb interactions between the particles. Instead it relies on the spatial constraints imposed by the short range interaction between the particles. We demonstrate the emergence of the FQHE phases in our model, by using an effective topological measure from graph theory, the Euler characteristic, applied in the real space of the system. This measure allows us to identify topological clustering mechanisms of the particles that lead to the FQHE phases. We study several cases, low filled, half-filled and densely filled systems with a variable number of particles.
Also we show that the FQHE phases in our model are robust to disorder. Our approach helps to gain a better understanding of the mechanisms leading to topological phases of matter in many-body systems like the FQHE phases, using simple models with minimal interaction rules, that could be experimentally realizable in cold-atomic systems and Bose-Einstein condensates.

For our calculations we use a Hubbard chain of spinless particles where only one particle is allowed per site,
\begin{equation}
H=U\sum_{i=1}^{M-1} n_{i}n_{i+1} + 
t\sum^{M-2}_{\substack{ \text{ i=1 } }}(c_{i}^{\dagger}c_{i+2} +h.c.)
\label{eq_h_1d}
\end{equation}
where $c^{\dagger}_{i},c_{i}$ are the creation 
and annihilation operators for a particle at site i in the chain and $n_{i}=c_{i}^{\dagger}c_{i}$ is the number operator
taking the value $0(1)$ for an empty(occupied) site.
The symbol U represents the interaction strength between the particles, while t is the value of the hopping between second nearest neighbouring sites in the Hubbard chain. 
The system consists of N particles distributed among M sites.
The filling of the system is $f=\frac{N}{M}$.
Eq. \ref{eq_h_1d} could describe hard-core bosons or spinless fermions.The calculations presented in this manuscript are for hard-core bosons which satisfy $[c_{i},c_{j}^{\dagger}] = (1-2n_{i})\delta_{ij}$, although we have performed calculations for fermions also. For our calculations we fix the value of the hopping at $t=1$. When two particles occupy nearest-neighboring (adjacent) sites in the chain they lift the total energy of the system by U. The nearest-neighbor repulsive interaction imposes a spatial constraint in the self-organization of the particles. For example when $t=0$ and  $f \le \frac{1}{2}$ the lowest Fock states of the system consist of cases where the particles are seperated by at least one empty site(hole) e.g. the minimum distance between the particles is 2 in units of the lattice constant\cite{1d,slava}. 
As we shall show these type of states survive even at the presence of the second nearest neighbor hopping in the chain (the second term in Eq. \ref{eq_h_1d}), as long as the interaction strength U is sufficiently large.

The Euler characteristic $\chi$ can be defined
as the difference between the number of vertices N
and the number of edges E between the vertices in a graph\cite{chen,oliver}
\begin{equation}
\chi=N-E
\label{euler}
\end{equation}
For our model, the particles act as vertices,
while edges are formed when the particles
occupy adjacent sites in the Hubbard chain.
Each pair of adjacent particles represents one edge which contributes energy U in the total energy of the system.
This way, the value of $\chi$ can be used to describe the clustering of the particles along the Hubbard chain\cite{2d,excited}.
For a system consisting of single particles, without
clustering, we have $\chi=N$. When all the particles condense
into one cluster of length N-1 then we have $\chi=N-(N-1)=1$.
The Euler counts the number of clusters contained in a single Fock state with a fixed configuration of particles\cite{excited}.
The clusters can be either single particles with length 0 or  lines of particles with maximum length N-1.
In general the system will lie in a superposition of various Fock states each one having its own configuration of particles and $\chi$. Then it is useful to define the $\chi$ of the superposition, by using the local curvature of the system at each site, as defined in graph theory for tree-like graphs \cite{chen,oliver}
\begin{equation}
K_{i}=\langle n_{i} \rangle - \frac{\langle d_i \rangle}{2}.
\label{eq_10}
\end{equation}
Here $\langle n_{i} \rangle$ is the occupation probability at
site i in the Hubbard chain, while $\langle d_i \rangle$ is the mean number of particles lying at the two adjacent sites around a particle at site i. The average in Eq. \ref{eq_10} is taken over the Fock states. For a single Fock state, we have $K_{i}=0$ for an empty site or for an occupied site with two adjacent particles($\langle n_{i} \rangle=1,\langle d_i \rangle=2$). For an occupied site with one adjacent particle($\langle n_{i} \rangle=1,\langle d_i \rangle=1$) we have $K_{i}=\frac{1}{2}$ and for an occupied site with no adjacent particles($\langle n_{i} \rangle=1,\langle d_i \rangle=0$) we have $K_{i}=1$. The Euler characteristic $\chi$ can be calculated by summing the curvature over all the sites of the Hubbard chain
\begin{equation}
\chi=\sum^{M}_{i=1}K_{i}.
\label{eq_13}
\end{equation}
This is analogous to the Gauss-Bonnet theorem of differential geometry, where a local curvature of a manifold is integrated over the whole surface of the manifold in order to get $\chi$. 

\begin{figure}
\begin{center}
\includegraphics[width=0.9\columnwidth,clip=true]{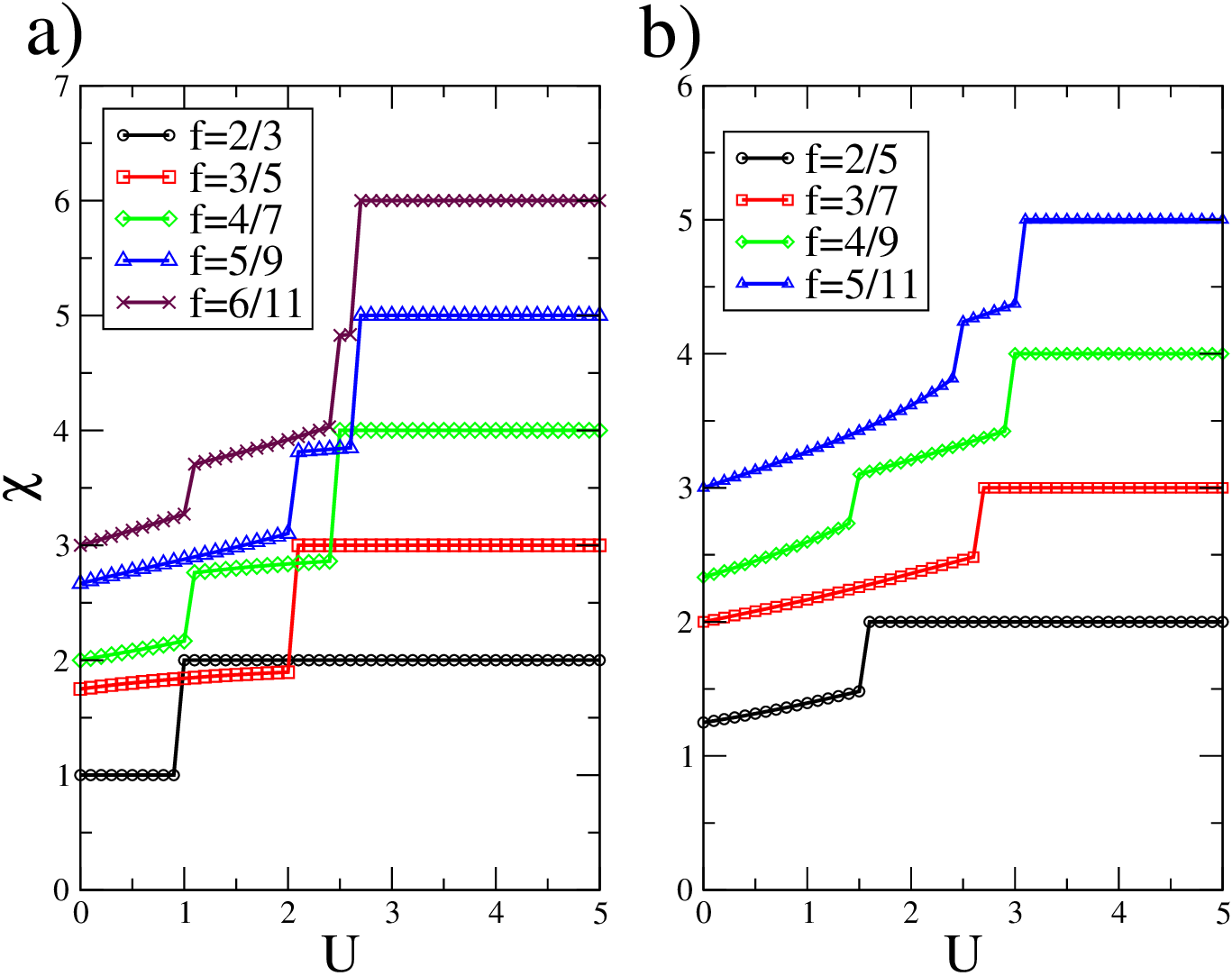}
\end{center}
\caption{a)The Euler characteristic $\chi$ for the ground state of system for odd denominator fillings and $M= 2N-1$, versus the strength of the interaction between the particles U. b)Some cases for $M > 2N-1$. In all cases the system reaches a topological quantum phase with $\chi=N$ for sufficiently strong U. The filling factors shown appear also in the FHQE leading to fractional conductance at quantized plateaus.}
\label{fig1}
\end{figure}

In Fig. \ref{fig1} we show several cases with $M \ge 2N-1 \rightarrow f \le \frac{N}{2N-1}$ for odd M corresponding to odd denominator fillings. For interaction strength U exceeding a critical value $U_c$ determined by the filling, a flat plateau at $\chi=N$ is formed, for all fillings studied, indicating the appearance of a topological phase. This phase consists of states that contain single particles seperated by a minimum distance 2 of the type $\ket{101 \dots 01000}$ (density-wave-order). These states appear also as the ground states of the system when $t=0$ in Eq. \ref{eq_h_1d}. No clustering/condensation of the particles is present in this case, since there is enough empty space in the system for the particles to be seperated from each other due to the nearest-neighbor repulsive interaction(the first term in Eq. \ref{eq_h_1d}), which minimizes the energy of the system. Below $U_{c}$ several non-flat plateaus at non-integer $\chi \ne N$ appear. These plateaus correspond to ground states which are superpositions of Fock states that contain a variable number of clusters. 

Essentially the repulsive interaction term in Eq. \ref{eq_h_1d} imposes a spatial constraint in the self-organization of the interacting particles. For sufficiently strong U and $f \le \frac{1}{2}$, the constraint manifests as a minimum distance 2 between the particles, as we have already pointed out. For $M = 2N-1 \rightarrow f=\frac{N}{2N-1}$ (Fig. \ref{fig1}a), the particles arrange in one classical state with density-wave-order, of the type $\ket{10101 \ldots 01}$, which has broken translational symmetry. This state lies at energy $E=0$ since there is no particle clustering and the second nearest hopping has no effect on the configuration of the particles, as they are spatially frozen. It is useful to examine a case with a small number of particles, for instance the $N=2,M=3$ case. The Hamiltonian written in the basis of all the possible Fock states $\ket{110},\ket{101},\ket{011}$,
is block diagonal containing two blocks. One block corresponds to the state $\ket{101}$ containing just a single zero element($H_1=0$), while another 2x2 block is formed for the states $\ket{110},\ket{011}$,
\begin{equation}
H_2=\left( \begin{array}{ccc}
U & t  \\
t & U  \\
\end{array} \right).
\label{equation1}
\end{equation}
This matrix has two eigenvalues $E_{1}=U+t$ and $E_{2}=U-t$.
Clearly when $U>t$ then $E_{2}>0$ and the ground state 
of the system becomes the Fock state with density-wave-order $\ket{101}$ lying at $E=0$. This state corresponds to the phase with $\chi=N$ reached at the critical interaction strength $U_{c}=t \Rightarrow U=1$ as shown in Fig. 1a.

For arbitrary N and M, by enumerating the sites in the Hubbard chain, we can see that the second nearest neighbor hopping allows a particle to hop only between sites of the same type, even or odd. This allows to write the Hamiltonian of the system in a block-diagonal form as we have explicitly shown for $N=2,M=3$. Two of the blocks always correspond to Fock states containing particles at even or odd occupied sites in the Hubbard chain. The rest of the blocks correspond to Fock states that contain particles at a fixed number of odd and even sites. The size of each of these blocks is determined by the number of permutations of the fixed odd occupied sites among the rest of the occupied even sites. The block diagonal structure of the Hubbard Hamiltonians allows us to diagonalize them more efficiently for large systems.

In Fig. \ref{fig1}b we show various cases with $M > 2N-1 \rightarrow f < \frac{N}{2N-1} \rightarrow f < \frac{1}{2}$ for odd M (odd denominator fillings). Again, the system reaches a quantum phase with $\chi=N$ for sufficiently strong U. We note the appearance of a topological phase for $f=\frac{1}{3}$ corresponding to $N=3,M=9$, although it is not shown in Fig. \ref{fig1}. Unlike the $M = 2N-1$ case, for $M > 2N-1$ the particles have enough spatial freedom to hop via the second nearest neighbor hopping, between the same type of sites in the Hubbard chain (even or odd). This leads to the formation of many possible Fock states with different particle configurations whose superposition acts as the ground state of the system. The second nearest neighbor hopping generates a dispersion around $E=0$ for these states, which results in a negative ground state energy of the system. In overall we see that the topological quantum phase with $\chi=N$ occurs at $E=0$ for $M = 2N-1$ and at $E<0$ for $M > 2N-1$. 

The mechanism leading to the topological phases in our model can be understood energetically by taking account of the block-diagonal form of the Hamiltonian. For odd denominator fillings and sufficiently strong U, the block that gives the lowest eigenvalue of the system, corresponds to states that contain single particles at odd sites of the Hubbard chain, seperated by a minimum distance 2. In contrast for even denominator fillings the Hamiltonian block which gives the lowest eigenvalue, always corresponds to states containing condensations/clusters of particles (either single particles or lines of particles) with a variable cluster number.
\begin{figure}
\begin{center}
\includegraphics[width=0.9\columnwidth,clip=true]{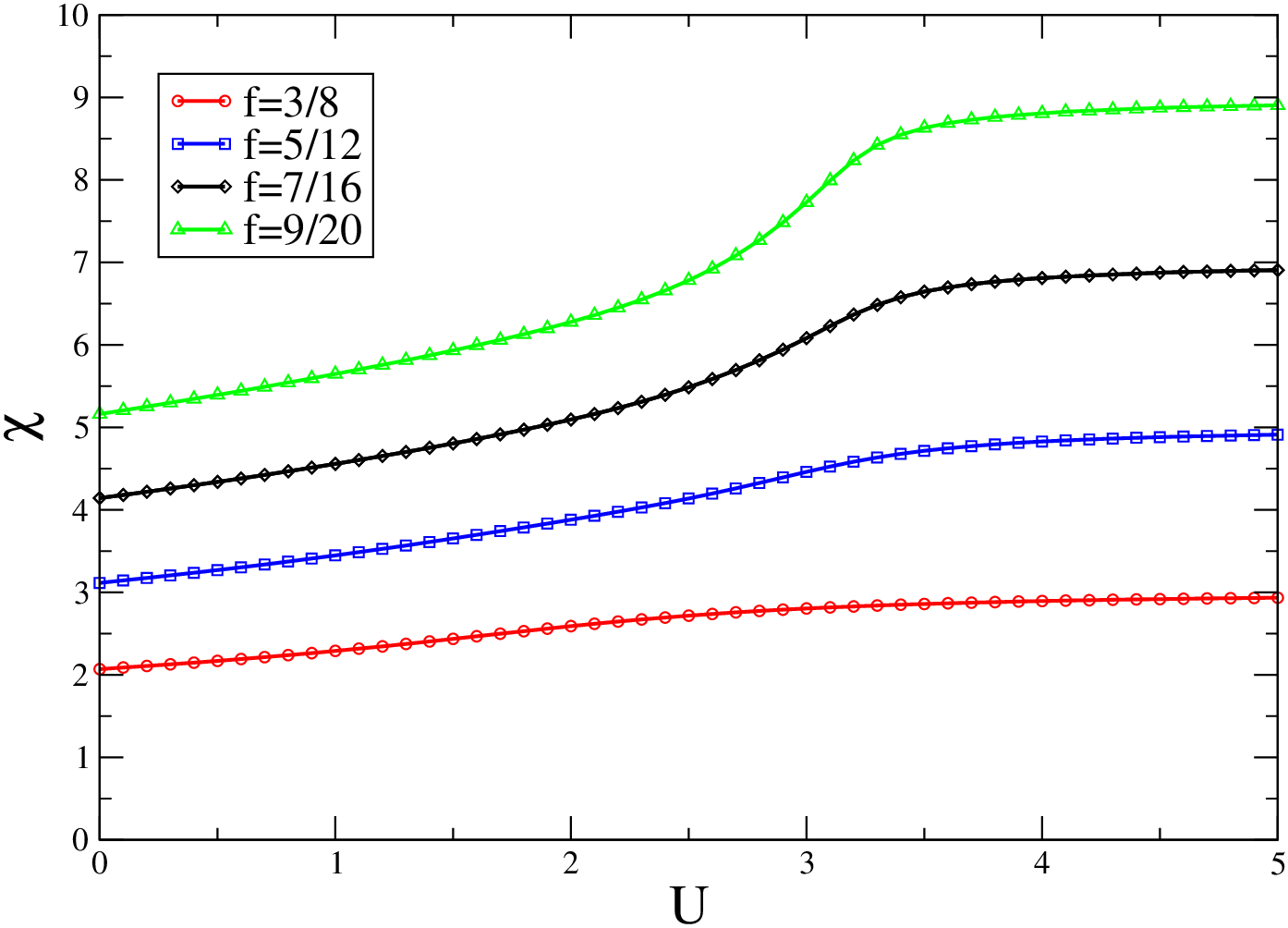}
\end{center}
\caption{The Euler characteristic $\chi$ for the ground state of the system for even denominator fillings and $M > 2N-1$ versus the strength of the interaction between the particles U. All cases approach asymptotically $\chi=N$ as U is increased. The topological phases with $\chi=N$ are not reached for any finite value of U unlike the odd denominator fillings. }
\label{fig2}
\end{figure}
Some cases for even denominator fillings are shown in Fig. \ref{fig2}. In contrast to the odd denominator fillings,  we do not observe a phase transition towards $\chi=N$. In this case the ground state of the system consists of states that contain a variable number of particle clusters leading to $\chi \ne N$. The value $\chi=N$ is approached only asymptotically as $U$ is increased since the limit $U \rightarrow \infty$ is equivalent to considering $t=0$ in Eq. \ref{eq_h_1d}.

\begin{figure}
\begin{center}
\includegraphics[width=0.9\columnwidth,clip=true]{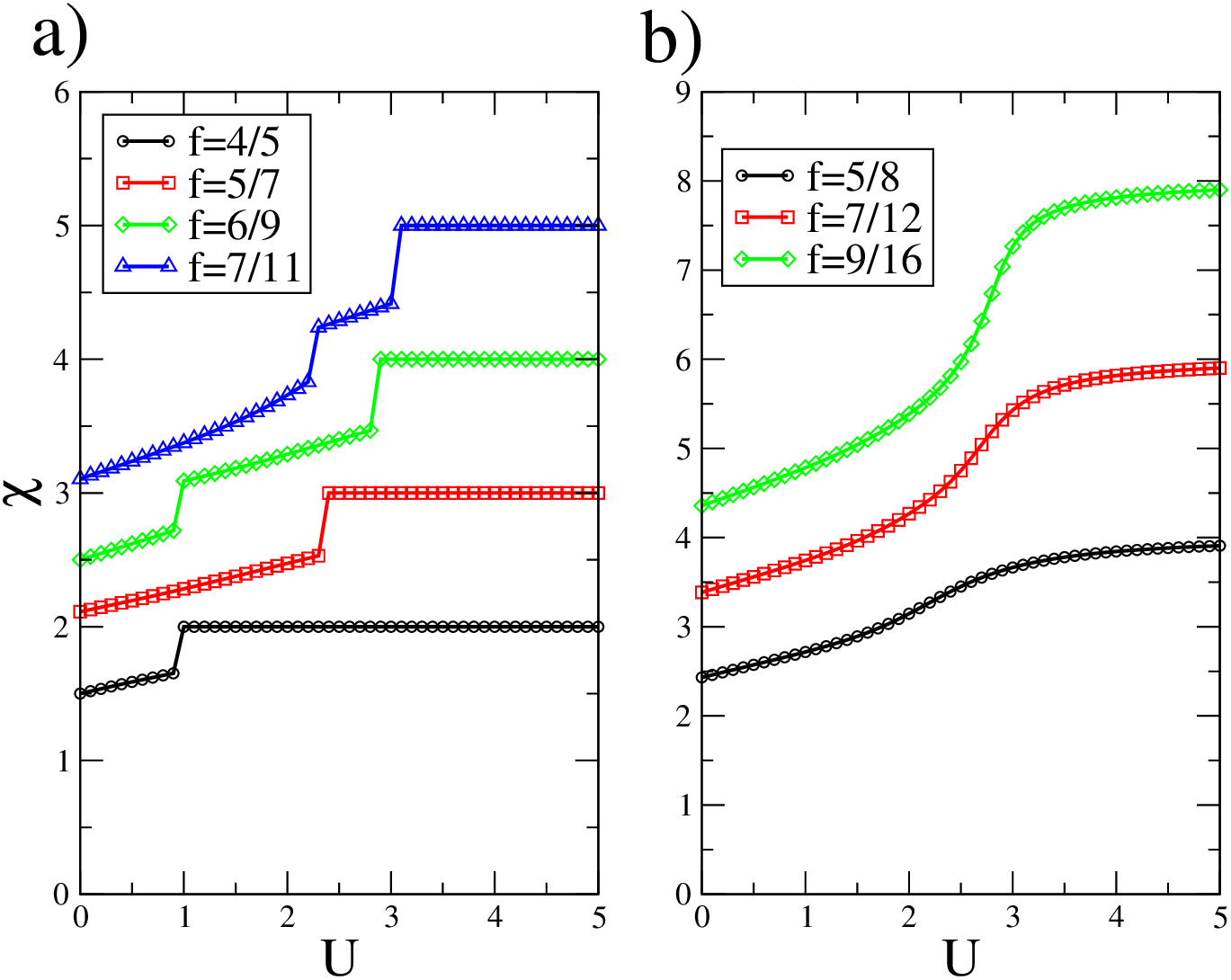}
\end{center}
\caption{a)The Euler characteristic $\chi$ for the ground state of system with odd denominator fillings for $M < 2N-1$ versus the interaction strength between the particles U. All cases reach a quantum phase with $\chi=M-N+1$ which is equal to the number of clusters contained in the ground state. The filling factors appear also in the FQHE. b)The respective cases for even denominator fillings approach $\chi=M-N+1$ only asymptotically as U is increased.}
\label{fig3}
\end{figure}

Another case of interest is when $M < 2N-1 \rightarrow f > \frac{1}{2}$ where the particles are spatially more restricted and cannot all stay seperated from each other. In Fig. \ref{fig3}a we show several examples
for this case and odd denominator fillings.
The system reaches a phase where $\chi$ is given
by the number of clusters in the ground state, instead of the number of particles. In general there exist $M-N+1$ clusters when $M < 2N-1$.  For example when $N=4,M=5$ the
phase with $\chi=2$ is determined by the states $\ket{11101},\ket{10111},\ket{11011}$
which contain two clusters. Again
the even denominator fillings shown in Fig. \ref{fig3}b do not reach this phase. For example the case $N=5,M=8$ approaches $\chi=2$ only asymptotically, since there are two clusters for $U \rightarrow \infty$ ($t=0$).

In overall, we can see that the topological phases in our model are reached for fillings $f=\frac{N}{M}$ where N and M have no common denominators and M is an odd integer. These are the filling factors where the fractionally quantized plateaus of the conductance occur in the FQHE\cite{Tsui,Stormer}.
We note that not all the odd denominator fractions with $f<1$ have been experimentally observed in a single system\cite{pan}.
The filling in our model is analogous to the filling factor in the FQHE, which represents the ratio of electrons to magnetic flux quanta. Essentially the M sites in our model correspond to the flux quanta that can be filled with N particles.

A fractionalization mechanism can be easily recognised in our model. As we have already analyzed, in a system with $M = 2N-1$, the particles in the ground state arrange in one classical state with density-wave-order of the type $\ket{10101 \ldots 01},$
due to the nearest-neighbor repulsive interaction. Removing any number of particles from this state will create holes that split in fractions, thus giving rise to fractional excitations in the system \cite{su,bergholtz}. These
type of states are responsible for the emergence of the $\chi=N$ phases in our model for the odd denominator fillings. The fractional excitations contained in these states can collectively carry current through the system leading to fractional conductance.

\begin{figure}
\begin{center}
\includegraphics[width=0.9\columnwidth,clip=true]{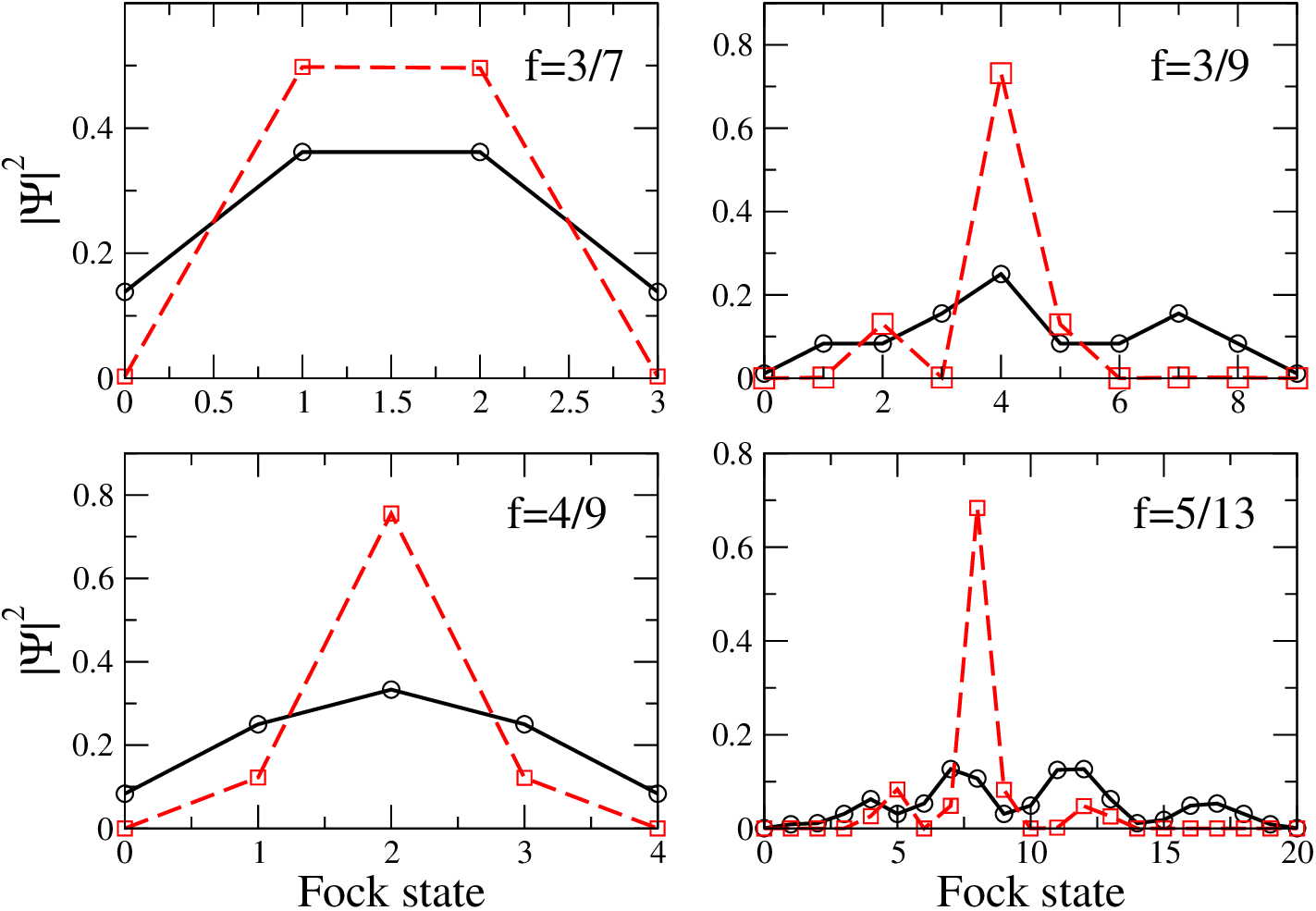}
\end{center}
\caption{The wavefunction probability for each Fock state
in the ground state of the system for different fillings.
Continuous black curves are the exact results obtained by diagonalizing the Hubbard Hamiltonian Eq. \ref{eq_h_1d} for $U=5,t=1$. The dashed red curves obtained using the Laughlin wavefunction Eq. \ref{eq_laughlin} catch qualitatively the trend of the exact results.}
\label{fig4}
\end{figure}

Not surprisingly we have found that the probability amplitudes for each Fock state in the ground state of the system can be qualitatively described by the Laughlin wavefunction adapted for the 1D case and for any filling f
\begin{equation}
\Psi =D \left[\prod_{N\ge i\ge j\ge1 }(\alpha(i-j))^{1/f} \right] \prod_{ i=1 }^{N}e^{-(\alpha i)^{2}}.
\label{eq_laughlin}
\end{equation}
where $i,j$ is the position of an occupied site in the Hubbard chain and D is a normalization factor.
Eq. \ref{eq_laughlin} gives the amplitude for one configuration of particles represented by the respective Fock state in our model. A comparison between the probability $|\Psi|^2 $ using Eq. \ref{eq_laughlin} with $\alpha=0.01$ and the exact result obtained by diagonalizing Eq. \ref{eq_h_1d} is shown in Fig. \ref{fig4}, for different fillings. Eq. \ref{eq_laughlin} catches the overall trend of the exact result, although
there are large quantitative differences for specific amplitudes.
 
Another feature of the FQHE that we are able to reproduce
with our model is the linear dependence of the gap from
the ground state to the first excited state, as a function of the filling.

Topology in our model manifests in the number of clusters contained in the Fock states, whose superposition is the ground-state wavefunction of the system. The Euler characteristic acts as a topological number characterizing the respective quantum phases. For odd denominator fillings
the system reaches a quantum phase with
$\chi=N$ for $M \ge 2N-1$ and $\chi=M-N+1$ for $M < 2N-1$.
In both cases $\chi$ counts the number of clusters which
are single particles for the $M \ge 2N-1$ case. The clusters act essentially as topological defects among the empty sites, embedded in an 1D manifold represented by the Hubbard chain. The Euler characteristic counts the number of these topological defects. In contrast the non-integer values of $\chi$ along the non-flat plateaus that appear for weak U below $U_c$, correspond to ground states which are superpositions of Fock states that contain a variable number of clusters, which cannot be used as a topological invariant.
\begin{figure}
\begin{center}
\includegraphics[width=0.9\columnwidth,clip=true]{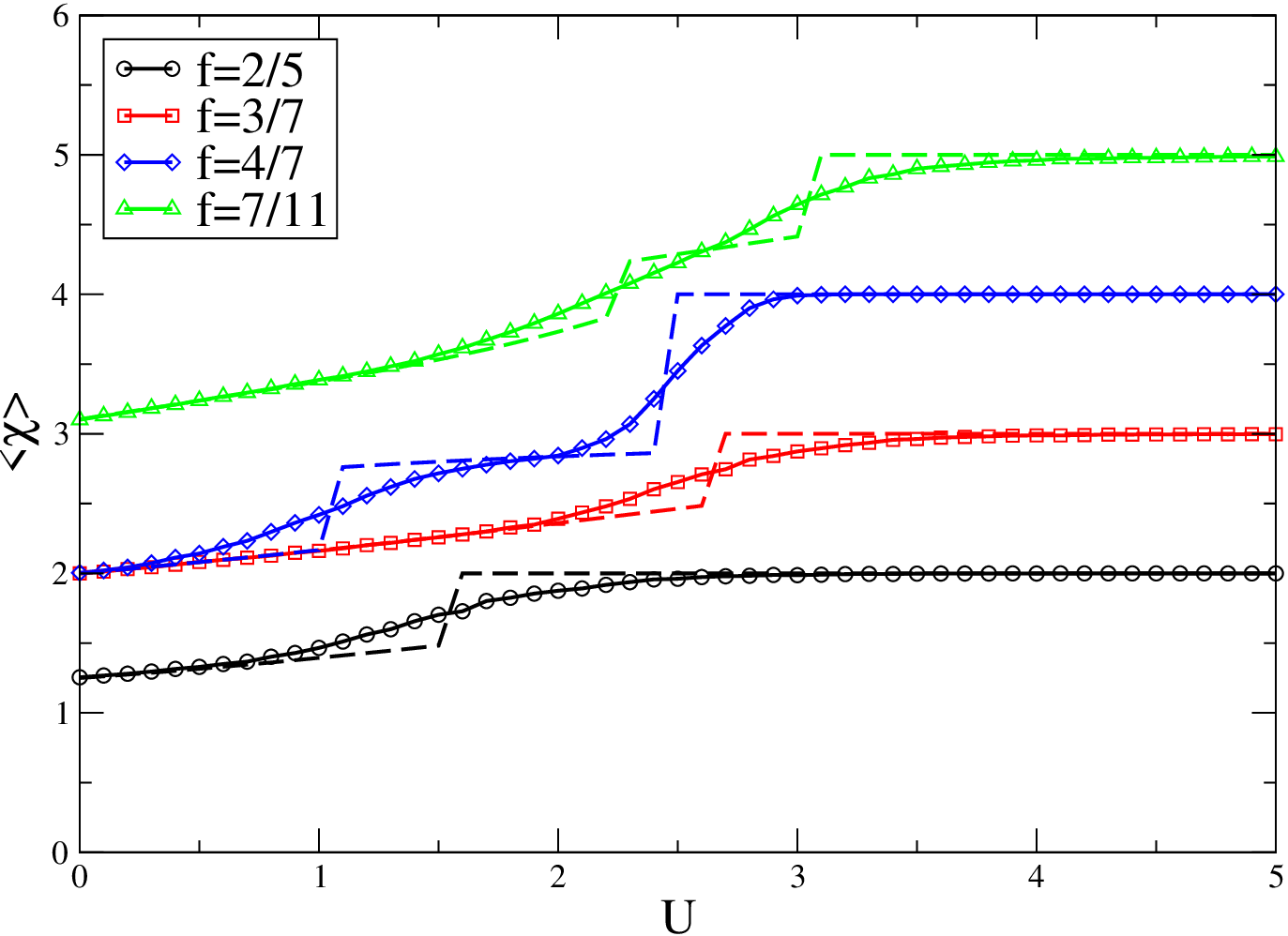}
\end{center}
\caption{The effect of disorder on the topological phases represented as continuous curves. The clean cases are shown with dashed lines for comparison. The phases at $\chi=N$ and $\chi=M-N+1$ are preserved for strong interaction strength U.}
\label{fig5}
\end{figure}

We remark that models similar to ours,
relying on repulsive interactions, have been proposed
in order to reproduce FQHE states and the respective wavefunctions in 1D \cite{bergholtz,guo,dyakonov}. Another similar Hubbard model has been used to describe transitions between superfluid, charge-density-wave(CDW) and supersolid phases in electron-phonon interacting systems and has been shown to contain a topological inequivalence between rings with even and odd number of sites \cite{gosh1,gosh2}. Also 1D FQHE setups have been studied experimentally \cite{timp}.

In Fig. \ref{fig5} we show the effect of disorder on the topological phases. The disorder is introduced by an on-site random potential term $H_d=\sum_{i=1}^{M} V_i n_{i}$. The random numbers $V_i$ are sampled from a box distribution
in the interval $[-W/2,W/2]$ for 1000 realizations of the disorder with $W=1$. The plateaus at $\chi=N$ and $\chi=M-N+1$ are preserved
despite the strong disorder. This robustness to
local perturbations is a typical feature of topological phases of matter. The plateaus at $\chi=N$ are also preserved if we add a small first nearest neighbor hopping in the Hubbard chain.

We note that although the topological FQHE phases presented in this manuscript are derived for hard-core bosons, we have verified their presence for fermions also.

To summarize, we have demonstrated the emergence of 1D quantum phases with topological characteristics,
at the filling fractions of the FQHE, in 
Hubbard chains with nearest neighbor interactions.
We have based our analysis on a topological measure
of graph theory, the Euler characteristic applied in real space of the system to describe the clustering of the interacting particles along the Hubbard chains. Our result shows
that the fractionalization and topological effects of the FQHE can be achieved in simple many-body systems that follow minimal interaction rules, lacking Coulomb interactions between the particles or external magnetic fields. This approach hints the universality of the FQHE, and how it could rely in general on spatial constraints imposed on the self-organization of particles in many-body systems. Apart from its fundamental significance our result could be useful in the realization of fractionalization and topological many-body mechanisms in cold atomic systems or Bose-Einstein condensates.

\section*{Acknowledgements}
We acknowledge resources and financial support provided by
the National Center for Theoretical Sciences of R.O.C. Taiwan and the Department of Physics of Ben-Gurion University of the Negev in Israel. Also we acknowledge support by the Project HPC-EUROPA3 (INFRAIA-2016-1-730897), funded by the EC Research Innovation Action under the H2020 Programme. In particular, we gratefully acknowledge the computer resources and technical support provided by ARIS-GRNET and the hospitality of the Department of Physics at the University of Ioannina in Greece.



\begin{thebibliography}{99}

\bibitem{spinhaldane}
F. D. M. Haldane,   Phys. Rev. Lett.  {\bf 1980}, 45, 1358.
 
\bibitem{haldane0}
 F. D. M. Haldane, Phys. Lett. A {\bf 1983a}, 93, 464.

\bibitem{AKLT}I. Affleck , T. Kennedy,  E.H.  Lieb and H. Tasaki,  Phys. Rev. Lett. {\bf 1987}, 59, 799.

\bibitem{Levin}  M. Levin  and  X.-G. Wen, 
  Phys. Rev. Lett. {\bf 2006} , 96, 110405.

\bibitem{Gu} X. Chen X,  Z.-C. Gu and   X.-G. Wen,  Phys. Rev. B {\bf 2010}, 82, 155138.

\bibitem{kitaev2}A. Kitaev A and J. Preskill,   Phys. Rev. Lett. {\bf 2006}, 96, 110404.

\bibitem{kitaev3}A. Y. Kitaev,   Ann. Phys. {\bf 2003}, 303, 2. 

\bibitem{alba}V. Alba, M. Fagotti and  P. Calabrese, J. Stat. Mech. {\bf 2009}, P10020.

\bibitem{alba1}V. Alba, M. Haque and M. Luchli, Phys. Rev. Lett. {\bf 2013}, 110 260403

\bibitem{hen} I. Hen  and M. Rigol, Phys. Rev. B {\bf  2009}, 80, 134508.

\bibitem{Hamma}A. Hamma,  R. Ionicioiu and P. Zanardi,  Phys. Rev. A {\bf 2005}, 71, 022315.

\bibitem{Calabrese}P. Calabrese and  A. Lefevre, Phys. Rev. A {\bf 2008}, f78, 032329.
 
\bibitem{Pollmann}F. Pollmann,  A. M. Turner,  E. Berg and M. Oshikawa,  Phys. Rev. B {\bf 2010}, 81, 064439.

\bibitem{amico}L. Amico,  R. Fazio, A. Osterloh and  V. Vedral,  Rev. Mod. Phys. {\bf 2008}, 80, 517. 

\bibitem{horodecki}R. Horodecki, P. Horodecki,  M. Horodecki M  and K. Horodecki, Rev. Mod. Phys. {\bf 2009}, 81, 865.

\bibitem{Tsui}D.C.  Tsui, H. L. Stormer,   and  A. C. Gossard, Phys. Rev. Lett. {\bf 1982}, 48, 1559.

\bibitem{Laughlin}E.B. Laughlin,   Phys. Rev. Lett. {\bf 1983}, 50, 1395.

\bibitem{Stormer}H. L. Stormer, D. C. Tsui and  A. C. Gossard, Rev. Mod. Phys. {\bf 1999}, 71, S298, S305.

\bibitem{Li}H. Li  and  F. D. M. Haldane,  Phys. Rev. Lett. {\bf 2008}, 101, 010504.

\bibitem{haldane_geometry} F. D. M. Haldane,  Phys. Rev. Lett. {\bf 2011}, 107, 116801. 

\bibitem{1d}I. Kleftogiannis and I. Amanatidis, Eur. Phys. J. B {\bf 2020}, 93, 84.  

\bibitem{slava}I. Kleftogiannis I,  I. Amanatidis and V. Popkov,  J. Stat. Mech. {\bf  2019}, 063102. 

\bibitem{chen}B. Chen  and G. Chen,  Gauss-Bonnet formula, finiteness condition, and asymptotic
characterization for graphs embedded in
surfaces Graphs, Combin. {\bf 2008}, 24, 159--183.

\bibitem{oliver}O. Knill,  A discrete Gauss-Bonnet type theorem, Elemente der Mathematik 67 {\bf  2012},1, pp1-44  arXiv 1009.2292 2010; 
O. Knill, A graph theoretical Gauss-Bonnet-Chern theorem, {\bf 2011}, arXiv 1111.5395.

\bibitem{2d}I. Kleftogiannis  and I. Amanatidis,  Eur. Phys. J. B {\bf 2019}, 92, 198. 

\bibitem{excited}I. Kleftogiannis  and  I. Amanatidis,  J. Stat. Mech. {\bf 2020},  083108.

\bibitem{pan}W. Pan, J. S. Xia, H. L .Stormer,  D. C. Tsui, C. Vicente, E. D. Adams, N. S. Sullivan, L. N. Pfeiffer, K. W. Baldwin and  K. W. West, Phys. Rev. B {\bf 2008}, 77, 075307.

\bibitem{su}W. P. Su  and  J. R. Schrieffer,   
Phys. Rev. Lett.  {\bf 1980}, 46, 738.

\bibitem{bergholtz} E. J. Bergholtz and  A. Karlhede, Phys. Rev. B {\bf 2008}, 77, 155308. 


\bibitem{guo}H. Guo,  S. Q. Shen  and S. Feng,    Phys. Rev. B {\bf 2012}, 86, 085124.  

\bibitem{dyakonov}M. I. Dyakonov,  IOP, Publishing, Journal of Physics: Conference Series {\bf 2013}, 456,  012008. 

\bibitem{gosh1}R. Pankaj and S. Yarlagadda, Phys. Rev. B {\bf 2012} 86, 035453.

\bibitem{gosh2}A. Ghosh and S. Yarlagadda, Phys. Rev. B  {\bf 2014} 90, 045140.

\bibitem{timp}G. Timp, R. Behringer, J. E. Cunningham and  R. E. Howard,  Phys. Rev. Lett. {\bf 1989}, 63, 2268.

\end{thebibliography}
\end{document}